\documentclass[aps,prb,twocolumn,showpacs,letterpaper]{revtex4-1}
\usepackage{amsmath}
\usepackage{ amssymb }
\usepackage{comment}
\usepackage{color}
\usepackage[usenames,dvipsnames,table]{xcolor}
\usepackage{graphics}\usepackage{epsfig}\usepackage{subfigure}
\usepackage{hyperref} 
\usepackage{multirow}
\usepackage{graphicx}
\usepackage{tabularx}
\usepackage{subfigure}
\begin{document}

\title{Charge density wave and spin $1/2$ insulating state in
  single layer 1T-NbS$_2$. }
\author{Cesare Tresca}
\author{Matteo Calandra}
\affiliation{Sorbonne Universit\'e, CNRS, Institut des  Nanosciences de Paris, UMR7588, F-75252, Paris, France}

\begin{abstract}

In bulk samples and few layer flakes, the transition metal dichalcogenides NbS$_2$ and NbSe$_2$
assume the H polytype structure with trigonal prismatic coordination of the Nb
atom. Recently, however, single and few layers of 1T-NbSe$_2$ with
octahedral coordination around the transition metal ion were
synthesized. Motivated by these experiments and 
by using first-principles 
calculations, we investigate the structural, electronic and dynamical properties of
single layer 1T-NbS$_2$. We find that single-layer 1T-NbS$_2$ undergoes a
$\sqrt{13}\times\sqrt{13}$ star-of-David charge density wave. Within
the generalized gradient approximation, the
weak interaction between the stars leads to an ultraflat band at
the Fermi level isolated from all other bands. The spin-polarized generalized gradient approximation
stabilizes a total spin $1/2$ magnetic state with opening of a $0.15$ eV band
gap and a $0.21\mu_B$ magnetic moment localized on the central Nb in the
star. Within GGA+U,  the magnetic moment on the central Nb is enhanced
to $0.41\mu_{B}$ and a larger gap occurs. Most important, this
approximation gives a small energy difference between the 1T and 1H polytypes
(only $+0.5$ mRy/Nb), suggesting that the 1T-polytype can be synthesized
in a similar way as done for single layer 1T-NbSe$_2$.
Finally we compute first and second nearest neighbors magnetic
inter-star exchange interactions 
finding $J_1$=9.5~K and $J_2$=0.4~K ferromagnetic coupling constants.

\end{abstract}

%
%
%
%
%
\maketitle

\section{Introduction}

\begin{figure}[!ht]
\centering
\includegraphics[width=\linewidth]{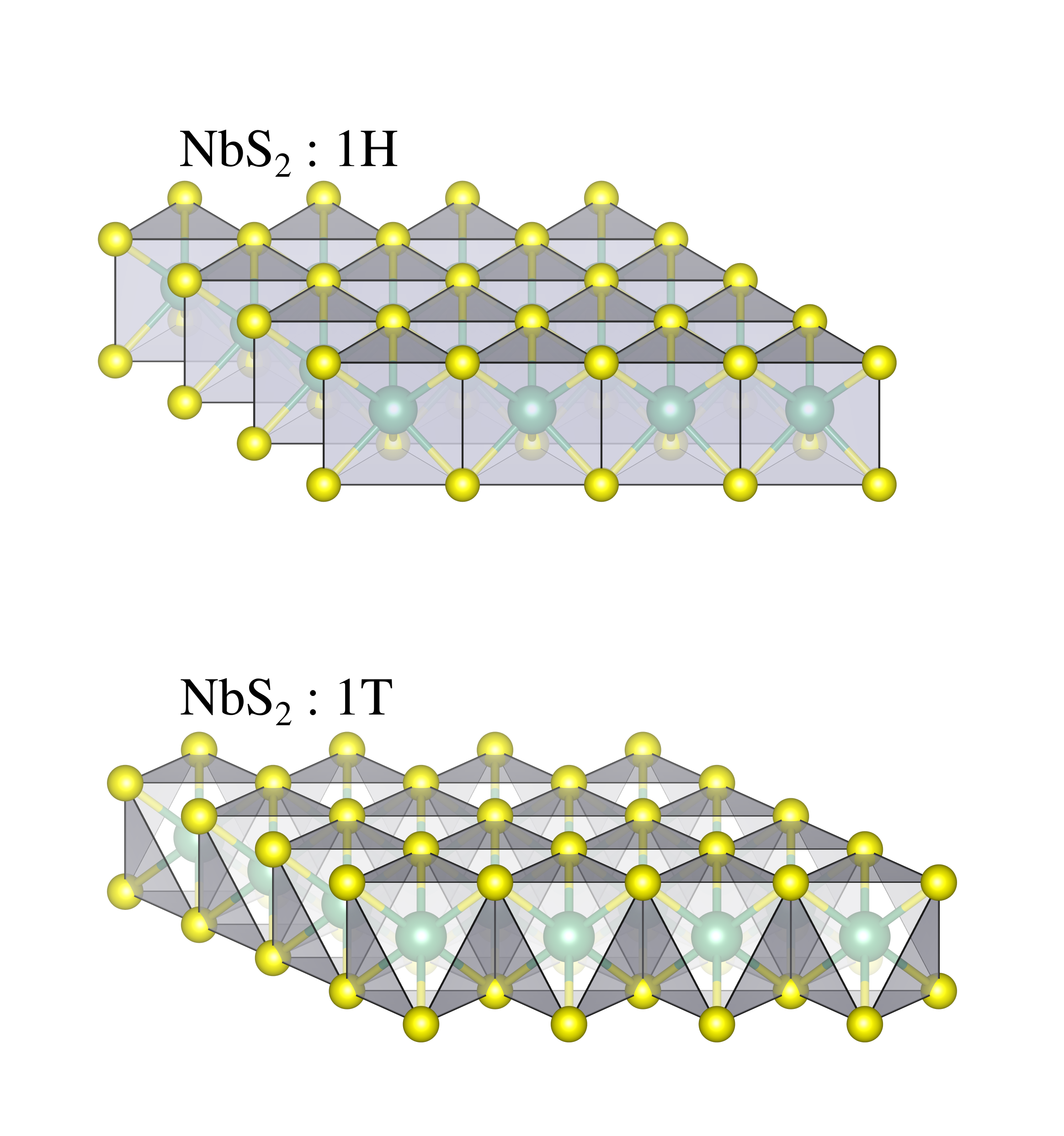}
\caption{Crystal structure of the 1H (top) and the 1T (bottom) NbS$_2$ single layer. The local coordination around a transition metal atom are
trigonal prismatic and octahedral respectively}\label{fig1}
\end{figure}

Bulk transition metal dichalcogenides (TMDs) of the form TCh$_2$, where T is a transition metal and Ch is a chalcogen
(Se, S, Te), are very versatile systems as their electronic and structural properties can be tuned not only by varying their
chemical composition but also by synthesizing different polytypes having the same chemical formula. 
The variation of the local coordination of the transition metal ion in different polytypes of a given TMD leads
to completely different physical properties\cite{DiSalvo}. For
example, 1T-TaS$_2$ with Ta in octahedral coordination, 
is a  correlated system which ground state is still very debated (Mott
insulator or correlated metal) \cite{FAZEKAS1980183,Fazekas2,PhysRevB.90.045134,Hoffman_kz}
 while 2H-TaS$_2$, with Ta in trigonal prismatic coordination, is a metal (the 1T and 1H polytypes are reported in Fig.\ref{fig1}).  
However, this tunability cannot be completely exploited as not all bulk TMDs can be synthesized in 1T and 2H polytypes either because the appropriate chemical and 
thermodynamical preparation conditions are unknown or because 
the energetic is unfavorable.
This is the case of bulk 1T-TiSe$_2$ that has never
been synthesized in the 2H polytype or, vice versa,  of bulk
2H-NbSe$_2$ and 2H-NbS$_2$ that crystallizes in the 2H polytype and
not in the 1T one,
although it has been reported\cite{B315782M} that 1T-NbS$_2$ bulk can be
synthesized under very special conditions.

Since different TCh$_2$ planes are bounded together by the weak van
der Waals interaction, it makes possible to isolate single layers of a large class of transition metal dichalcogenides \cite{Novoselov2D,ColemanLiquid} (a single layer here refers to a trilayer TCh$_2$ unit).
The exfoliation of bulk TMDs into bi-dimensional (2D)
crystals, beside being interesting in itself as it allows to
investigate a variety of phenomena in low dimension, paves the way to  different synthesis techniques, unfitted for bulk systems but feasible in 
few layers flakes. An example is the phase transition between the hexagonal and monoclinic phases of monolayer MoTe$_2$
achieved by electrostatic doping\cite{MoTe2FET} or the transition between the 2H and 1T and 1T$^{\prime}$ phases obtained by liquid
exfoliation\cite{ColemanLiquid,GokiEda}. 

More recently it has been shown that the 1T-NbSe$_2$ polytype
can be stabilized either as a single layer on top of bilayer graphene kept at 500-590~$^\circ$C during epitaxy\cite{Nakata2016}
or by applying a pulsed local field through the STM tip 
at the surface of bulk 2H-NbSe$_2$\cite{doi:10.1021/acs.chemmater.7b03061}.

The physical properties of single layer 1T-NbSe$_2$ turned out to be completely different from that of single layer 1H-NbSe$_2$ as the former
is a spin $1/2$ Mott-Jahn Teller insulator undergoing a $\sqrt{13}\times\sqrt{13}$ charge density wave\cite{Nakata2016, PhysRevLett.121.026401}, 
while the latter is a metal undergoing a $3\times3$ charge density wave\cite{MakNbSe2,Ugeda,CalandraMazin}. 
Most important, it has been recently shown\cite{PhysRevLett.121.026401} that density functional theory calculations (DFT) with local LDA/GGA kernels do not
explain the stabilization of the  1T-NbSe$_2$ single layer phase with respect to the 2H one, as this transition occurs via a 
correlated mechanism involving vibrations and the stabilization of a magnetic state that can be addressed within the DFT+U approximation.
Thus, given the broad perspectives offered by these new synthesis techniques, theoretical calculations can be used
to spot new TMD phase that can be experimentally accessed and to describe their structural and electronic properties.

Bulk 2H-NbS$_2$ is isoelectronic and isostructural to 2H-NbSe$_2$,
however it stands somewhat at odd with respect to other transition metal dichalcogenides as it displays no charge density wave (CDW) at low temperature\cite{Leroux2HNbS2}. On the contrary, when NbS$_2$ single-layer is grown on Nitrogen-doped 6H-SiC(0001) terminated with single or bilayer graphene, 
it adopts the 1H-NbS$_2$ polytype and STM images show a $3\times3$
reconstruction\cite{LinCDW1HNbS2}, but if 1H-NbS$_2$ is
grown epitaxially on Au(111) no charge density wave is detected\cite{NbS2gold}. Given the different properties of
NbS$_2$ in the 2D limit, it is natural to investigate the possible stability of
other polytypes and the formation of magnetic and charge density wave phases.

In this work, by using density functional theory calculations, we investigate the possible synthesis of single layer 1T-NbS$_2$
together with its structural, vibrational and electronic properties. We study the stability with respect to the
single layer phase and we calculate magnetic couplings.

\section{Computational details}

Density functional theory calculations are performed using the \textsc{Quantum-Espresso} code~\cite{QEcode,QE-2017}.
For Nb (Ta) we use ultra-soft pseudopotentials from Vanderbilt
distributions\cite{PhysRevB.41.7892} including semicore
states and two projectors for $s$ and $p$ channels and valence
configuration $4s^2$, $4p^6$, $4d^5$, $5s^1$ ($5s^2$, $6s^2$, $5p^6$,
$5d^3$). 
For  S (Se) we use norm-conserving pseudopotentials with empty
$d$-states in valence and the following valence configuration $3s^2$, $3p^3$, $3d^0$ ($4s^2$, $4p^3$, $4d^0$).

We use an energy cutoff up to 45~Ry (540~Ry for the charge density) for all the calculations. For the exchange correlation energy we take the generalized gradient approximation (GGA) and the GGA+U one. 
 
The charge density integration over the Brillouin Zone (BZ) is performed using an uniform $20\times 20\times 1$  Monkhorst and Pack grid~\cite{PhysRevB.13.5188} for the 1T and 2H-polytypes ($6\times 6\times 1$ and $5\times5\times1$ for the $\sqrt{13}\times\sqrt{13}$ and $4\times4$ CDW phases respectively) and a $0.01$~Ry Gaussian smearing. For the total energy comparison among magnetic solutions of the $\sqrt{13}\times\sqrt{13}$ reconstruction we reduce the smearing to $0.0001$~Ry increasing the BZ grid to $12\times 12\times 1$ (for the evaluation of exchange constants we use super-cells: we scale the BZ sampling grid to assure the same density used in the other calculations). The surface is simulated by considering a supercell with about $10$~\AA~of vacuum along the c-axis between the periodic images. We use the theoretical in-plane lattice parameters and perform full structural optimization of the internal degrees of freedom.
Phonon modes in the undistorted 1T-phase are calculated in linear response theory\cite{QEcode,QE-2017} over 19 phonon wave-vector mesh in the irreducible BZ using an uniform $20\times 20\times 1$ reciprocal space mesh for sampling the electronic states.

\section{Results and discussion}

\subsection{High-symmetry 1T-NbS$_2$ structure.}

We start by performing geometrical optimization of the undistorted 1T-NbS$_2$ structure (3 atoms/cell). For completeness and to achieve a better
understanding of the transition metal/chalcogen hybridization, we
also
calculate the theoretical GGA structural parameters and electronic
structures of 1T-NbSe$_2$, 1T-TaS$_2$ and 1T-TaSe$_2$  high symmetry
phases.  
We obtain the lattice parameter and internal coordinates reported in Tab.~\ref{tab0}. 
As it can be seen the sulfur dichalcogenides are somewhat compressed in the basal plane with respect to Se dichalcogenides. 
The smaller in-plane parameter is accompanied by a smaller chalcogen height ($h_{Ch}$) and a smaller
tetragonal distortion of the octahedral crystal symmetry around the transition metal ion.
\begin{table*}[]
\begin{center}
\begin{tabular}{ l l c c c c }
\hline
\hline
\multirow{5}*{\includegraphics[scale=0.75]{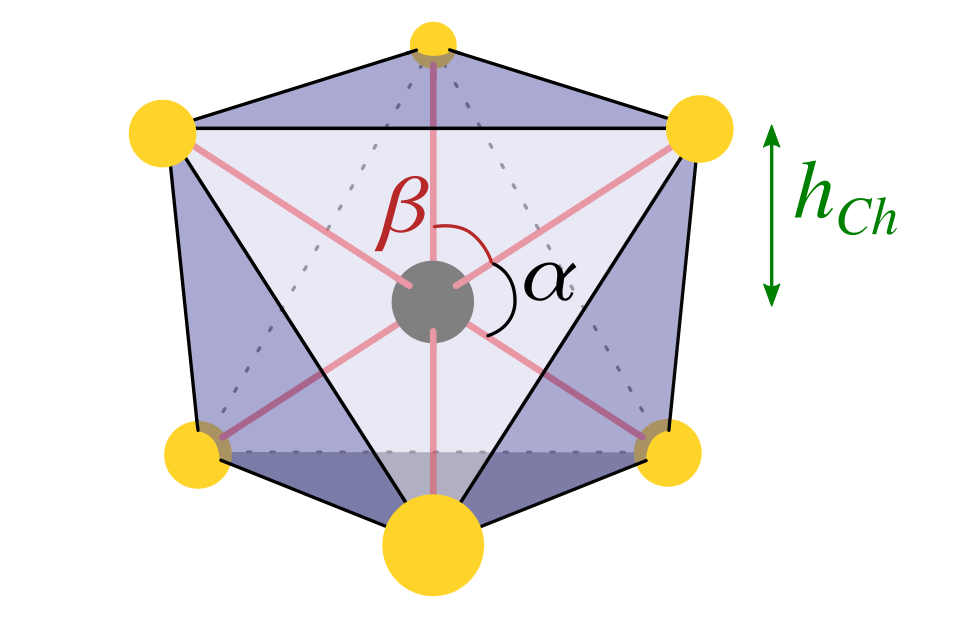}} & System               & ~$a$ (\AA) & $h_{Ch}$ (\AA) & $\alpha$ (deg) & $\beta$ (deg)   \\
\cline{2-6}
&~NbS$_2$         &  ~~3.35$^{~~}$   & 1.55  &  94.7  &  85.3  \\
&~NbSe$_2$        &  ~~3.48\cite{PhysRevLett.121.026401}   & 1.69  & 97.5    & 82.5   \\
&~TaS$_2$         &  ~~3.38$^{~~}$  & 1.53  &  94.0  &  86.0  \\
&~TaSe$_2$        &  ~~3.50$^{~~}$   & 1.65  & 95.7  &  84.3 \\
\\
\\
\\
\\
\\
\hline
\hline
\end{tabular}
\end{center}
\caption{Theoretical internal coordinates for 1T-TCh$_2$ systems. In
  the inset image we report in gray the T atoms and in yellow the Ch
  one, the quantities tabulated are highlighted.}
\label{tab0}
\end{table*}
\begin{figure*}[]
\centering 
\subfigure[~1T-NbS$_2$ \{$d_{z^2-r^2}(\Gamma)=$47\%\}]%
{\includegraphics[width=0.45\linewidth]{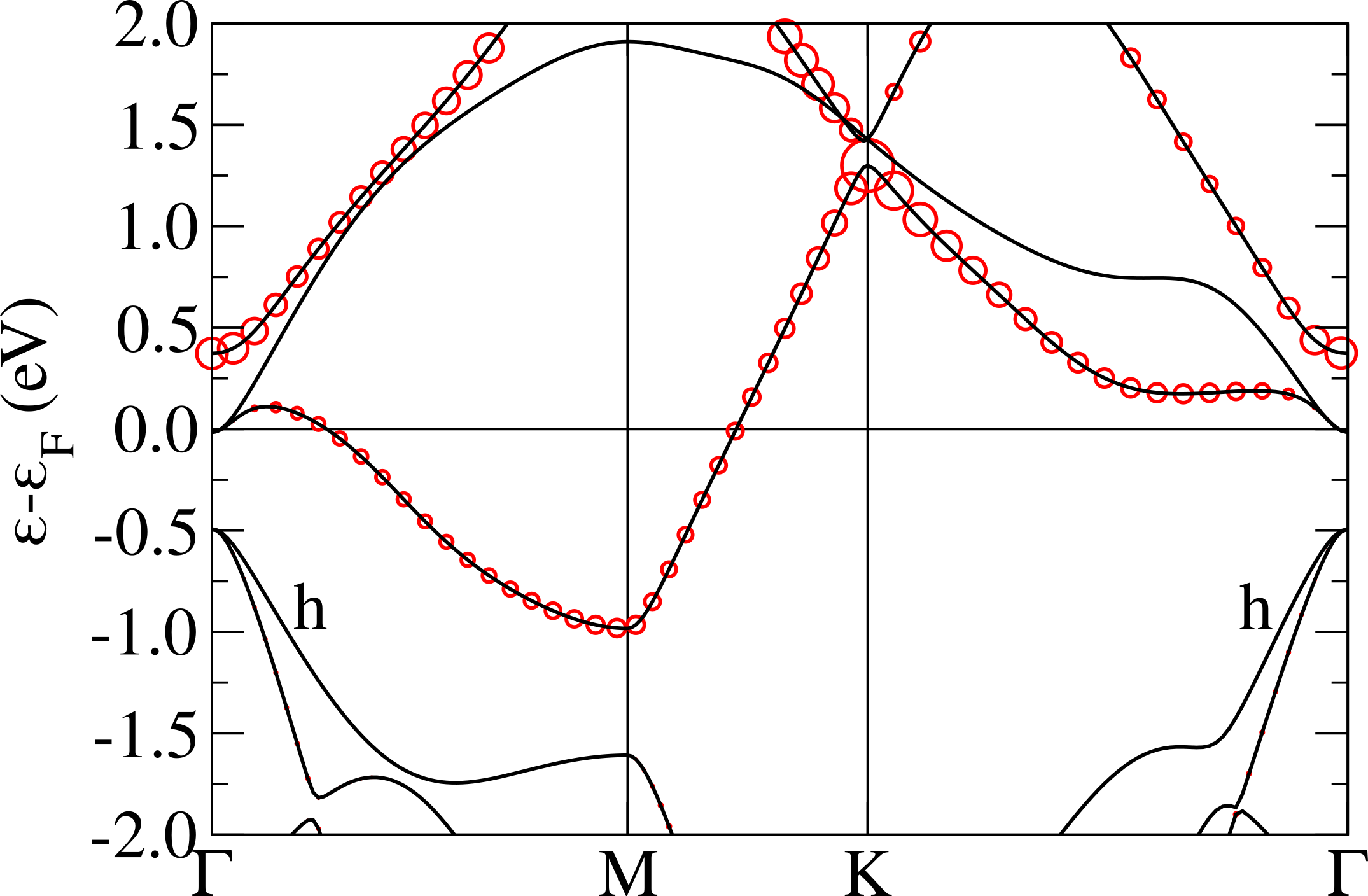}}\qquad%
\subfigure[~1T-NbSe$_2$ \{$d_{z^2-r^2}(\Gamma)=$54\%\}]%
{\includegraphics[width=0.45\linewidth]{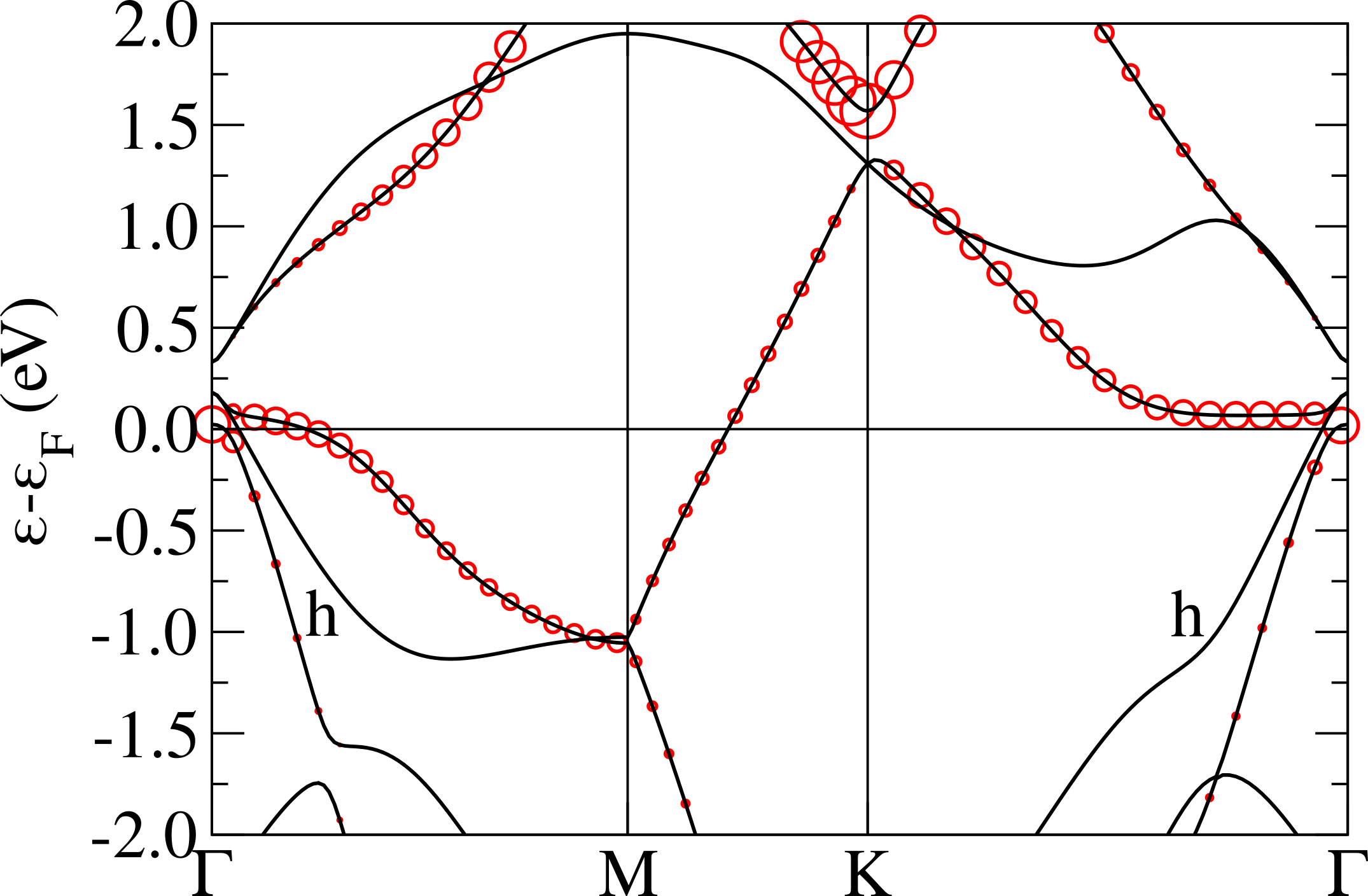}}\\
\subfigure[~1T-TaS$_2$ \{$d_{z^2-r^2}(\Gamma)=$46\%\}]%
{\includegraphics[width=0.45\linewidth]{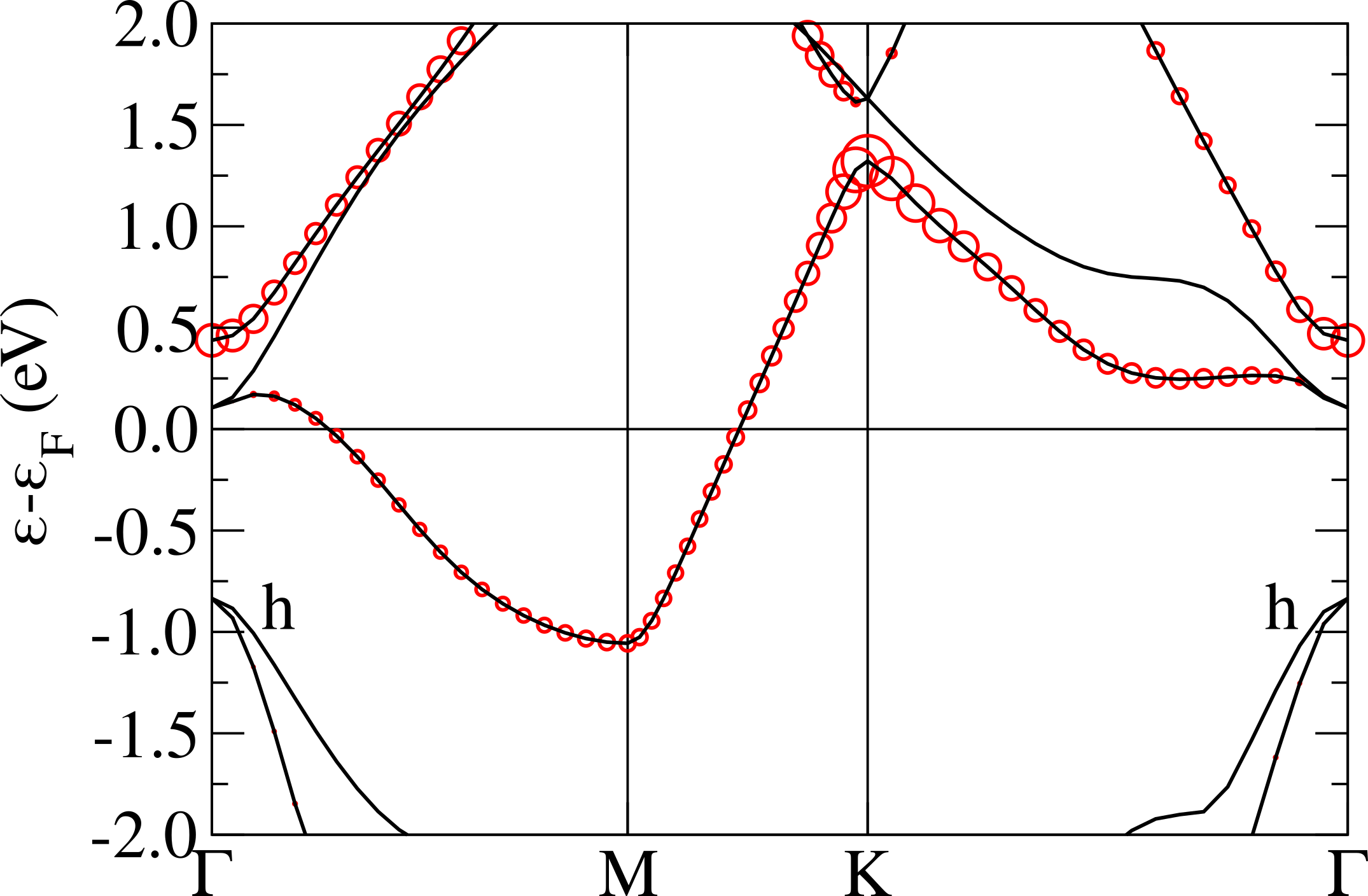}}\qquad%
\subfigure[~1T-TaSe$_2$ \{$d_{z^2-r^2}(\Gamma)=$54\%\}]%
{\includegraphics[width=0.45\linewidth]{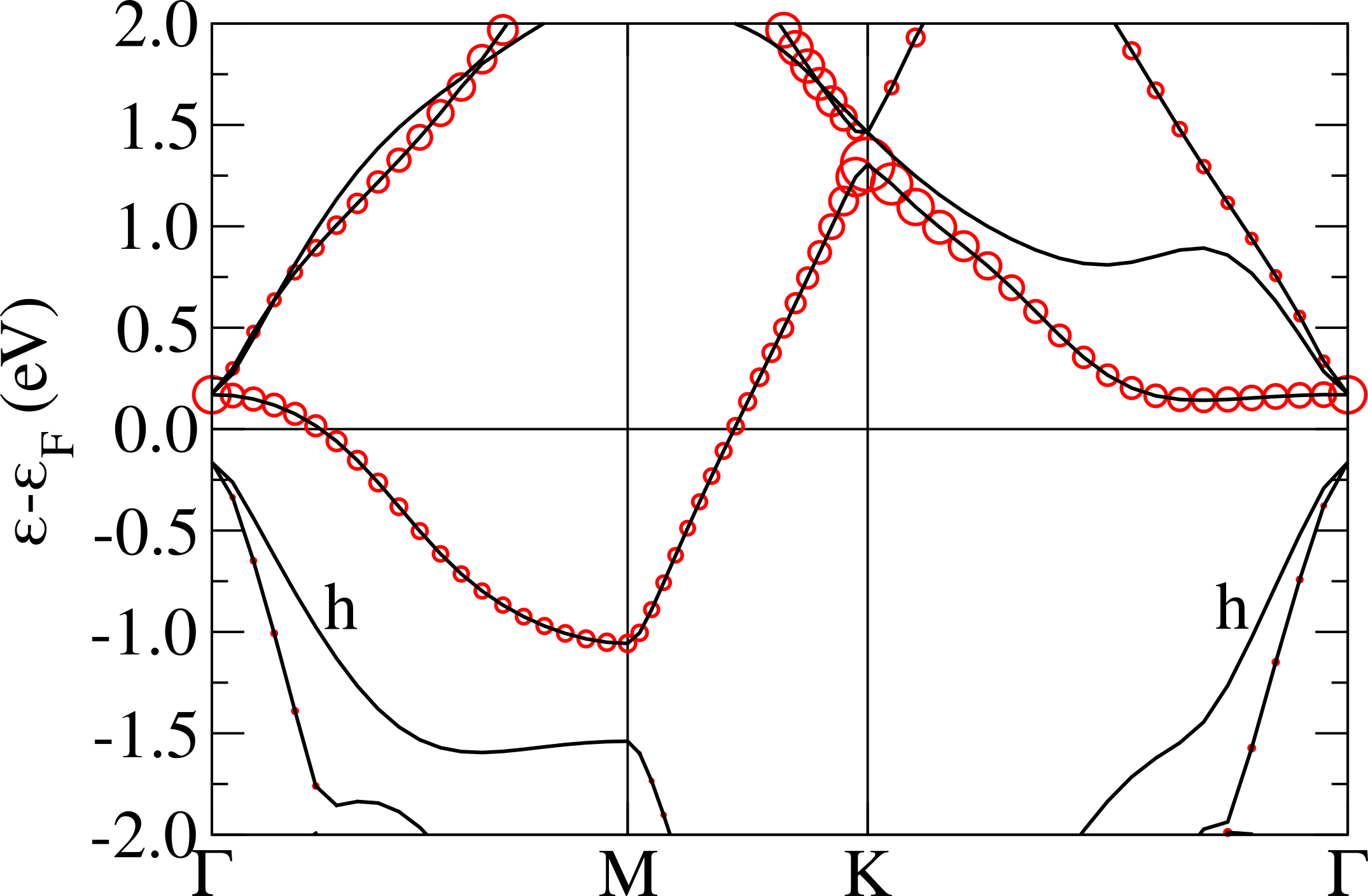}}
\caption{Band structures for TCh$_2$ compounds (T=Nb,Ta; Ch=S,Se) in
  the 1T-polytype. The size of the circles is proportional to the
  Nb$_{d_{z^2-r^2}}$ character of the eigenvalues, the percentage of the
  $d_{z^2-r^2}$ component at $\Gamma$ for each system is reported in the
  subcaption.}\label{fig2}
\end{figure*}%

This is relevant as the electronic structures of all these highly symmetric polytypes  are similar but with important differences that can be in part attributed
to the amount of Jahn-Teller trigonal distortion of the octahedral crystal field around the transition metal and 
in part to the alignment of the chalcogen and transition metal levels\cite{Canadell,Mattheiss}.

In more details, the octahedral crystal field splitting leads to  triply degenerate t$_{2g}$ orbitals ($d_{x^2-y^2}$, $d_{z^2-r^2}$, 
$d_{xy}$) and doubly degenerate e$_g$ orbitals ($d_{xy}$,
$d_{xz}$) at higher energy (we adopt here the same convention of Ref.~\cite{Mattheiss} for the crystal axes). 
The trigonal distortion of the octahedron is identified
by the bond angles centered at the transition metal ion and having bonds to the nearest chalcogens
(see picture in Tab. \ref{tab0}). In an undistorted octahedron  $\alpha=\beta=90^{o}$, while in the present
cases there is a substantial deviation from the ideal values.

The crystal field for a trigonal distorted octahedron splits the  
t$_{2g}$ orbitals in a twofold degenerate state ($d_{x^2-y^2}$, $d_{xy}$) and a single degenerate $d_{z^2-r^2}$
state. We label this energy separation at zone center ``apparent Jahn-Teller splitting''. 
In the case of 1T-NbS$_2$ and 1T-TaS$_2$, the apparent trigonal Jahn-Teller splitting at the {\bf $\Gamma$}
point is positive (namely the band originating from the $d_{z^2-r^2}$ state is higher in energy with respect to the
twofold degenerate one arising from the $d_{x^2-y^2}$ and $d_{xy}$ atomic orbitals) and very similar in 
magnitude for both systems,  as it can be seen in Fig.\ref{fig2}.
Surprisingly, in selenides, despite a  larger distortion, the apparent
Jahn-Teller splitting is almost zero or negative.

This apparent contradiction can be solved by considering the hybridization between the 
transition metal t$_{2g}$ bands and the other occupied chalcogen bands
at zone center (labeled ``h'' in Fig.\ref{fig2}). In sulfides,
this band is mainly formed by sulfur 3$p$ states.
In  TaS$_2$ this separation is larger than in NbS$_2$,
mainly due to the larger energy misalignment between the sulfur 3$p$  and the Nb 4$d$ or Ta 5$d$ states. 
The situation is very different in selenides, where the hybridization between the Se $p$ states and the 
Ta or Nb $d$ states is strong (stronger in Nb than in Ta) and leads to completely counterintuitive results
with respect to crystal field theory. For example, the larger octahedral distortion occurs in NbSe$_2$, but here we find an apparent
negative Jahn-Teller splitting. Finally, in TaSe$_2$ the
apparent Jahn-Teller splitting is almost zero as the crystal field and the hybridization perfectly cancels out and the t$_{2g}$ bands
become almost threefold degenerate at zone center. Furthermore the top of what were the chalcogen $p$-bands in sulfides
becomes mixed with $d$-states in selenides (particularly evident in NbSe$_2$).

The different magnitude of the hybridization explains why in sulfides one expect t$_{2g}$ manifolds separated
by the chalcogen states while in selenides the character is more entangled \cite{PhysRevLett.121.026401}.

Finally, it is worth mentioning that as the 1T-polytype breaks the inversion symmetry, 
we investigate the magnitude of relativistic effects in 1T-NbS$_2$ finding them negligible, as expected given the relatively light atoms involved.

Having understood the electronic structure of the highly symmetric phase in comparison with other 1T compounds,
we compare the energy of single layer 1T-NbS$_2$ with the 1H-NbS$_2$ polytype. We find that
1H  is more stable by 
approximately $7.2$ mRy/Nb (see also Tab. \ref{tab1}), similarly to what happens for the NbSe$_2$ case\cite{PhysRevLett.95.237002,0953-8984-30-32-325601,Pasquier1TNbS2}.
This large energy difference prevents an highly symmetric 1T phase to form in experiments.
\begin{table}[]
\begin{center}
\begin{tabular}{ l c c c c }
\hline 
\hline 
 System               & $\Delta E$ & $\Delta E_U$ & $a$ & $a_U$   \\
\hline 
~1H         &  ~0.0   & ~0.0  &  3.346  &  3.326  \\
~1T         &  +7.2   & +4.2  &  3.360  &  3.357  \\
~4$\times$4 &  +5.8*   & +1.9  &  13.485*  & 13.428   \\
~$\sqrt{13}\times\sqrt{13} $        &  +4.3   & +1.2  & 12.200  & 12.123  \\
~$\sqrt{13}\times\sqrt{13}$ (FM)  &  +4.4   & +0.5  & 12.198  & 12.126  \\
\hline 
\hline 
\end{tabular}
\end{center}
\caption{Calculated energy difference among different polytypes and
  CDW phases. The energy differences are in mRy/Nb. In the last two
  columns the theoretical equilibrium lattice constants with and
  without $U$ ($a$ and $a_U$ respectively) are reported in \AA,
  $U=2.87$eV for all the calculations. 
The experimental in-plane lattice parameter for bulk 2H-NbS$_2$ is
$3.31$\AA\cite{Jellinek}.
 The (*) means we have two different CDWs, practically degenerate in energy.}
\label{tab1}
\end{table}

\begin{figure*}[]
\centering 
\includegraphics[width=0.93\columnwidth]{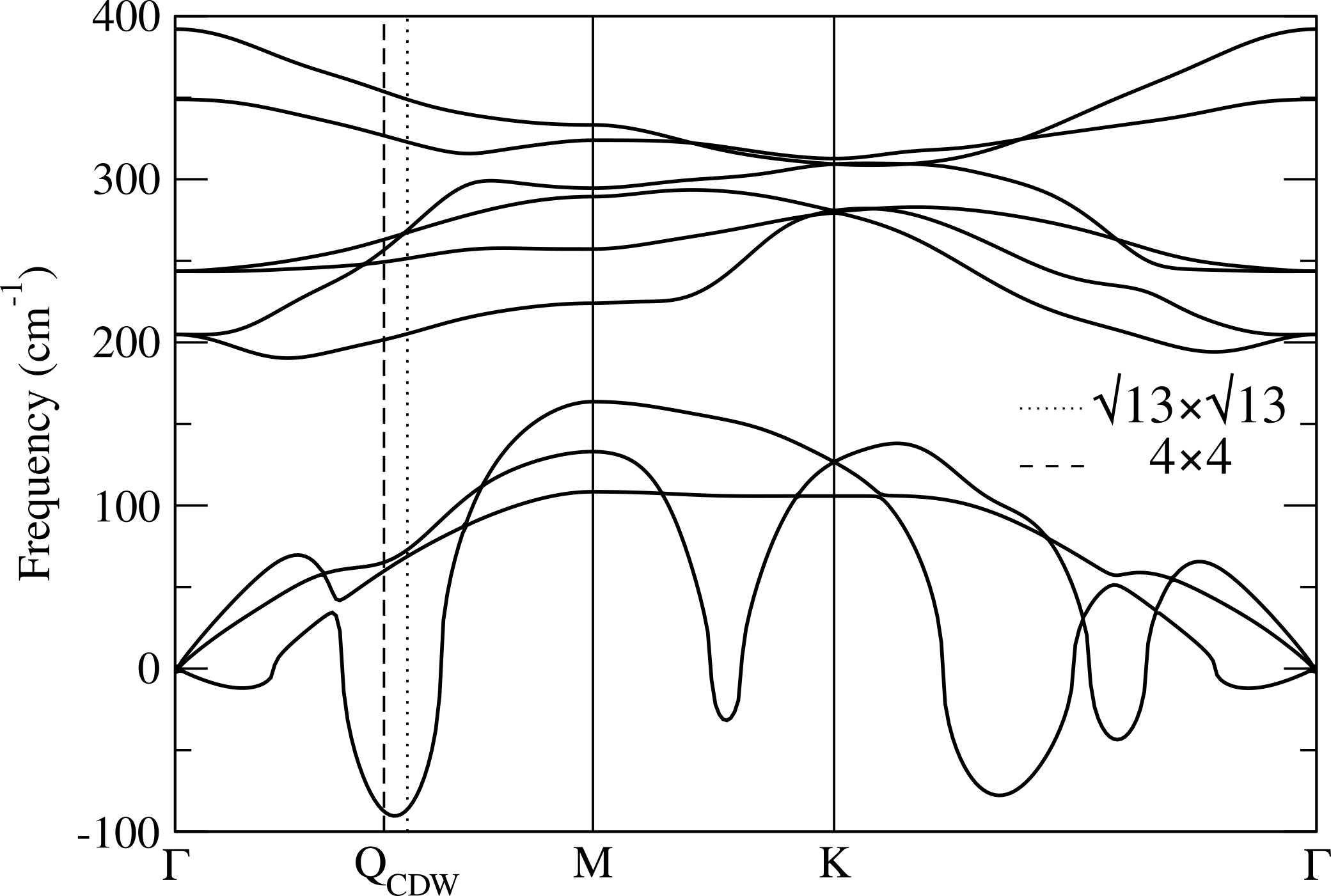}
\hspace{0.750cm}\includegraphics[width=0.9\columnwidth]{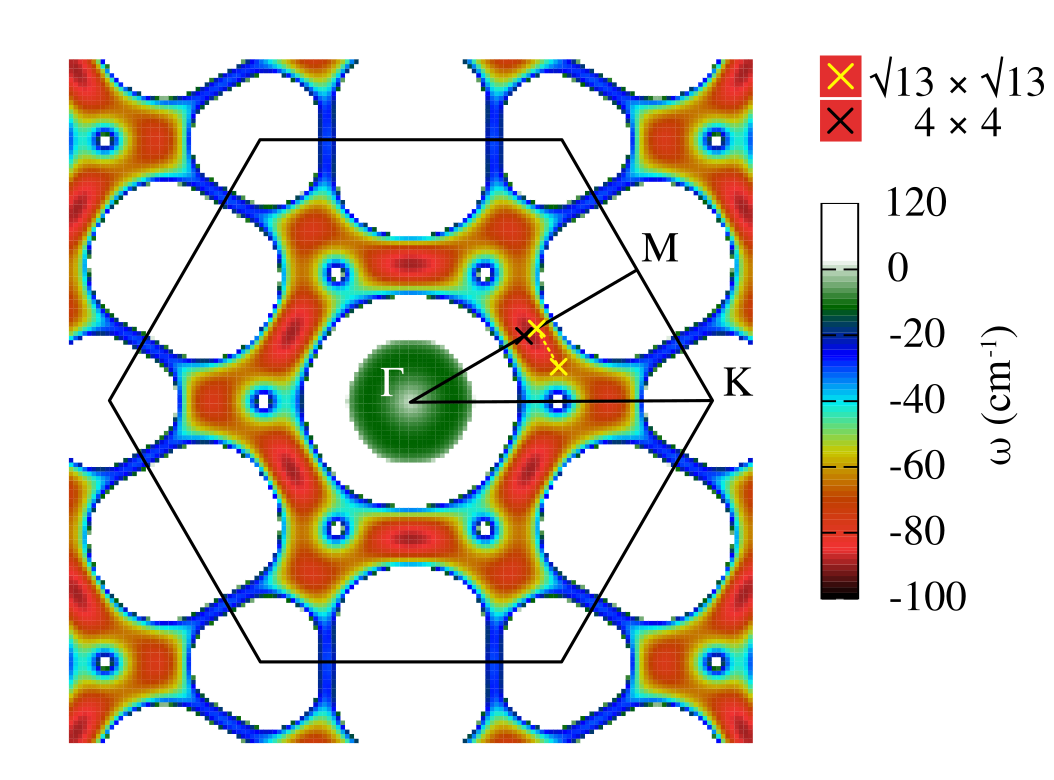}
\caption{Left: phonon dispersion along high-symmetry directions of the high-symmetry 1T-NbS$_2$ phase. 
The projections of the ordering vectors related to the $4\times 4$ and to the $\sqrt{13}\times\sqrt{13}$ CDWs onto the $\Gamma$-M line 
are marked with vertical lines. Right: Distribution of negative phononic frequencies in the BZ for undistorted 1T-NbS$_2$.  The wave vectors belonging to the $4\times 4$ and to the $\sqrt{13}\times\sqrt{13}$ CDWs are marked with crosses of different colors}\label{fig3}
\end{figure*}%

In order to inspect for possible CDW instabilities, we then calculate the phonon dispersion of single layer 1T-NbS$_2$.
As shown in Fig.\ref{fig3}  we found strongly unstable phonon modes.
To better identify the wavevector of the most unstable phonon
frequencies in the BZ, we also perform a 2D plot of the instability in Fig.\ref{fig3}.
At the harmonic level we find that the two most likely instabilities have wavevector compatible with a $\sqrt{13}\times\sqrt{13}$\cite{PhysRevB.82.155133} and a $4\times4$ CDWs.

\subsection{Charge density wave phases}

\begin{figure*}[]
\centering
\includegraphics[width=5.5cm]{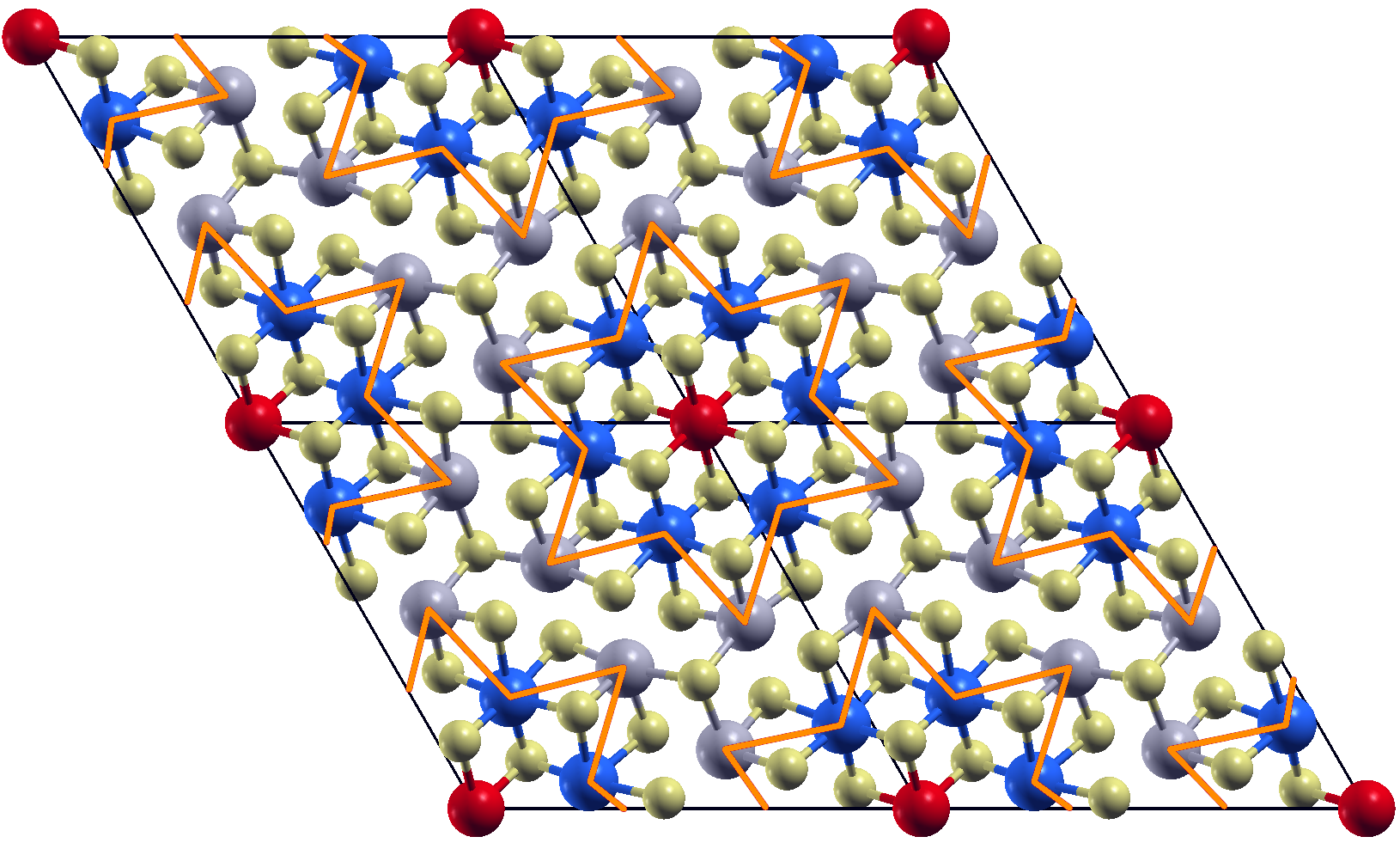}\includegraphics[width=5.5cm]{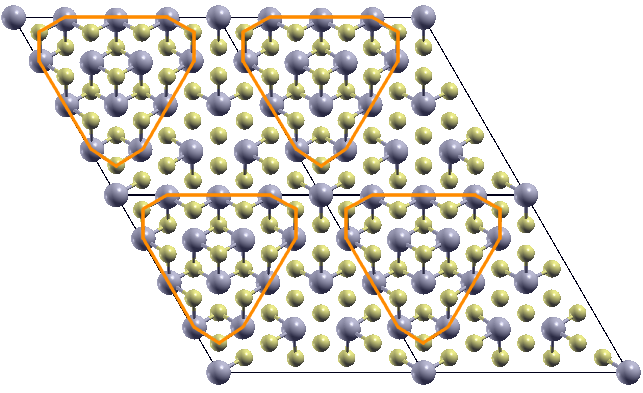}\includegraphics[width=5.5cm]{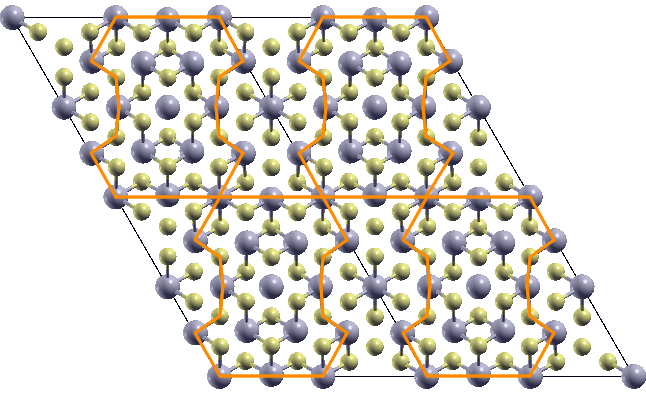}
\caption{Left: Optimized crystal structure for the $\sqrt{13}\times\sqrt{13}$ CDW phase. The tree nonequivalent Nb sites are highlighted: central Nb in the star (red), Nb belonging to the peripheral atoms of the $\sqrt{7}\times\sqrt{7}$ cluster (blue) and the other Nb atoms are in gray. S atoms are shown in yellow. Center and right: Optimized crystal structure for the $4\times4$ CDW. The orange lines are a guide for the eye to better recognize the building blocs of each reconstruction.}\label{fig4}
\end{figure*}

We  perform geometrical optimization within the GGA approximation in $4\times4$ and $\sqrt{13}\times\sqrt{13}$ supercells starting from initial configurations
obtained by displacing the atomic coordinates following the patterns
of the most unstable phonon modes. 
In both case, we find structures that are substantially more stable
than the highly symmetric ones. In the case of a $4\times4$
supercell, we find two different reconstructions that are practically degenerate
in energy (see Fig.\ref{fig4}). Both $4\times4$ CDW, however,
seems to try to form some kind of star-of-David reconstruction, but
the non ideal periodicity hinders the complete formation.
This is confirmed by the fact that $\sqrt{13}\times\sqrt{13}$ is the
most stable  reconstruction, 
with an  energy gain of $\approx 2.9$~mRy/Nb with respect to the highly symmetric
1T-NbS$_2$ phase, however still with an
energy loss of
$\approx 4.3$~mRy/Nb with respect to the highly symmetric 1H-NbS$_2$ phase.

The optimized $\sqrt{13}\times\sqrt{13}$ structure is shown in
Fig.\ref{fig4} (we also report the Wyckoff positions in App.~A),
 and it results dynamically stable (see App.~B), the relative energy differences among the different phases considered are reported in Tab.~\ref{tab1}.

\begin{figure}[]
\centering 
\includegraphics[width=0.9\linewidth]{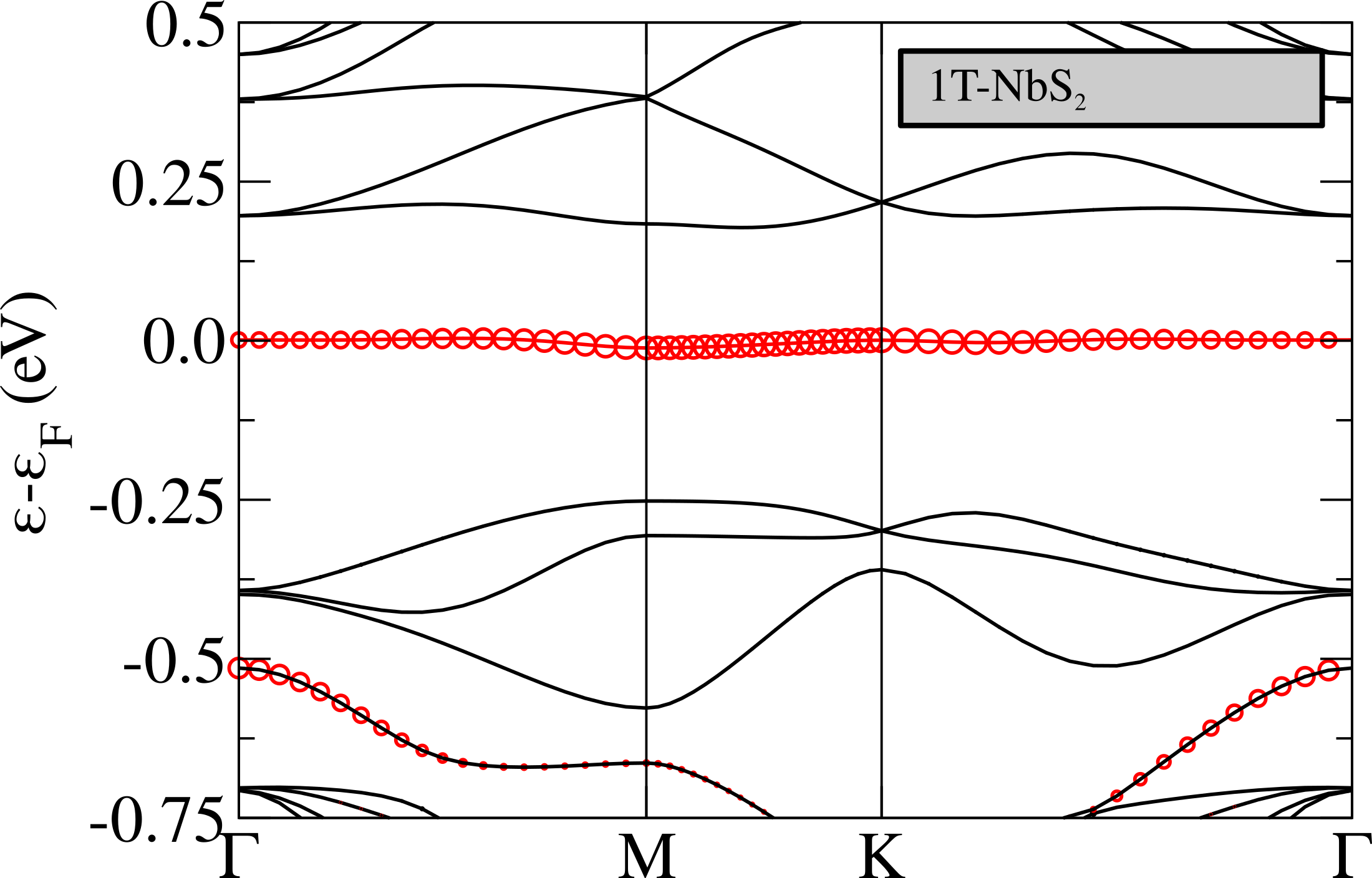}\\\vspace{0.25cm}
\includegraphics[width=0.9\linewidth]{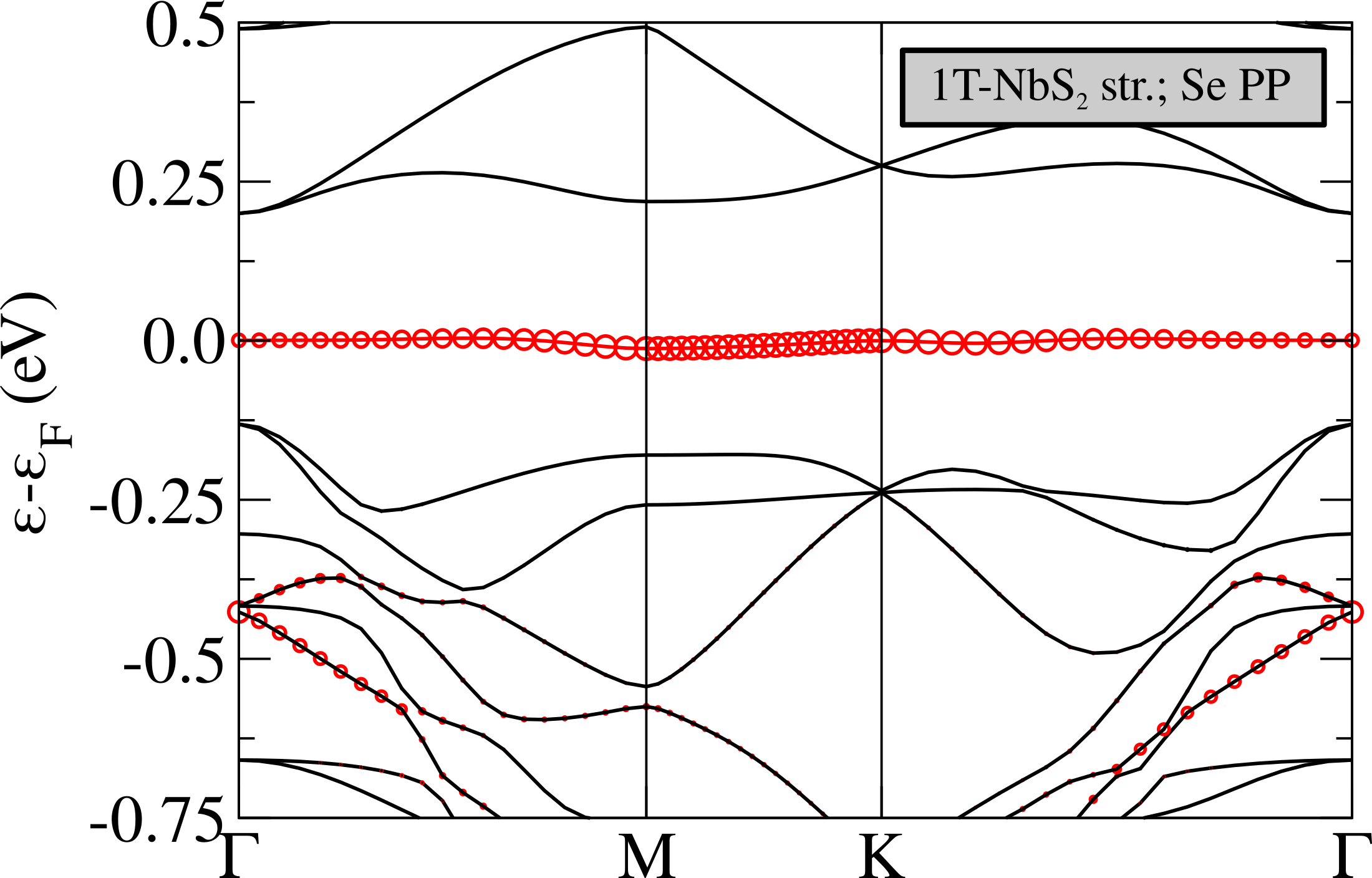}\\\vspace{0.25cm}
\includegraphics[width=0.9\linewidth]{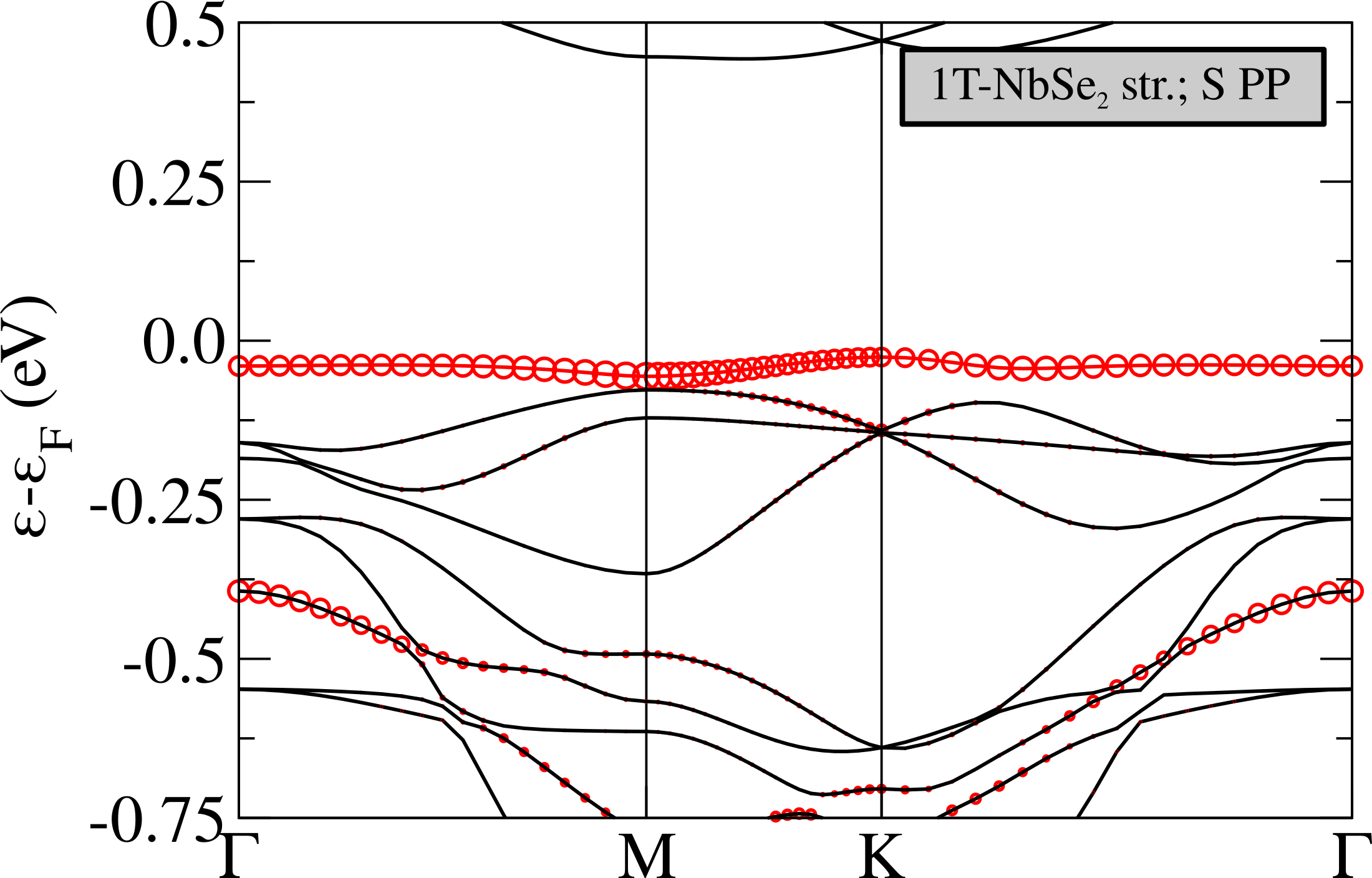}
\caption{Electronic structure of 1T-NbS$_2$ in the $\sqrt{13}\times\sqrt{13}$ CDW phase 
  (top panel) and of 1T-NbS$_2$ in the CDW phase where the 
  pseudopotential of S has been replaced by the one of Se (central panel). 
 Electronic structure of 1T-NbS$_2$ using lattice 
  parameters and internal coordinates of NbSe$_2$ $\sqrt{13}\times\sqrt{13}$ in the 
  CDW phase (bottom panel). 
The size of the circles is proportional to the $d_{z^2-r^2}$
character of the central Nb in the star  (the percentage of the 
central Nb $d_{z^2}$ component at $\Gamma$ at the Fermi level are 11\% (top), 13\% (central) and 9.6\% (bottom))}\label{fig5}
\end{figure}%

The non-magnetic electronic structure in the $\sqrt{13}\times\sqrt{13}$ phase is shown in Fig.\ref{fig5}~(top). It is characterized by the presence of an extremely flat band at the Fermi level having a non-negligible $d_{z^2-r^2}$ character related to the central Nb atom in the star. The flat band is isolated from the others and is located in the middle of the gap, this is in analogy with what happens in 1T-TaS$_2$\cite{PhysRevB.90.045134,PhysRevB.97.045133,PhysRevB.93.214109,PhysRevB.82.155133,PhysRevB.79.220515} and in contrast with the 1T-NbSe$_2$
case\cite{PhysRevLett.121.026401,Pasquier1TNbS2,0953-8984-30-32-325601}
where the flat band is entangled with chalcogen states.

In order to disentangle the effects of chemistry and distortion in
determining the energy position of the flat band with respect to the
chalcogen states, we calculate the
non-magnetic electronic dispersions of 
(i)  NbS$_2$ CDW structure in which we substitute the S atoms with Se keeping, however, the structure unchanged (labeled ``1T-NbS$_2$
str.; Se PP'' in Fig.\ref{fig5}),
(ii) NbS$_2$ using the crystal structure of 
1T-NbSe$_2$ in the CDW phase (labeled ``1T-NbSe$_2$ str.; S PP'' in
Fig.\ref{fig5}). 
These calculations should be directly compared with the case of NbSe$_2$
reported in  Ref.\cite{PhysRevLett.121.026401} (Fig.3, left panel) where,
at the GGA level, the flat band lies in the middle of Se states and is
not isolated from the others. This comes mostly from a 0.25~eV upshift of the 
lower occupied states at zone center.

Calculation (i) allows us to determine the effect of alignment
between Se/S states with Nb ones. As it can be seen the effect of
replacing the S with a Se pseudopotential is an up-shift of the Se states
at zone center, in agreement with what happens in the ideal
high-symmetry undistorted 1T-NbS$_2$/1T-NbSe$_2$ phases.  
However this up-shift is still not large enough to mix the flat band
with the other occupied bands, as it happens in the
$\sqrt{13}\times\sqrt{13}$ phase of
1T-NbSe$_2$\cite{PhysRevLett.121.026401}. 
If, on the contrary, we use the NbSe$_2$ $\sqrt{13}\times\sqrt{13}$
structure with the S pseudopotential, as in calculation (ii), we see
that the results is to up-shift mostly the occupied states very close
to the flat band. However, this is not what happens in the 
$\sqrt{13}\times\sqrt{13}$ phase of 1T-NbSe$_2$ as in this system, at the GGA level, the
top of the Se states at $\Gamma$ are empty and are at
higher energies then the flat band.
It follows that the effect is not properly chemical neither
structural, but it is a cooperative effect of the two.
This aspect is a general feature strictly related to the chalcogen
atom involved in the compound. In fact the same behavior is
observable also in TaS$_2$ 
 and TaSe$_2$ 
 (see Refs.\cite{PhysRevLett.121.026401,PhysRevB.90.045134,PhysRevB.97.045133}). 

As shown in Fig.\ref{fig5}~(top), the Nb$_{d_{z^2-r^2}}$ band is
extremely flat, with a dispersion of $\sim 0.016$~eV. 
This implies a small Fermi velocity, a low kinetic energy and an high
peak in the density of the states at the Fermi level. It is then
natural to expect electronic instabilities to occur.
We then perform spin-polarized calculations stabilizing an insulating
ferrimagnetic solution with an energy loss of about 0.1~mRy/Nb with respect to the metallic non-magnetic solution (see Tab. \ref{tab1}). 
Thus, even in the absence of an Hubbard term, GGA stabilizes a magnetic
state.

\begin{figure*}[]
\centering 
\includegraphics[width=\columnwidth]{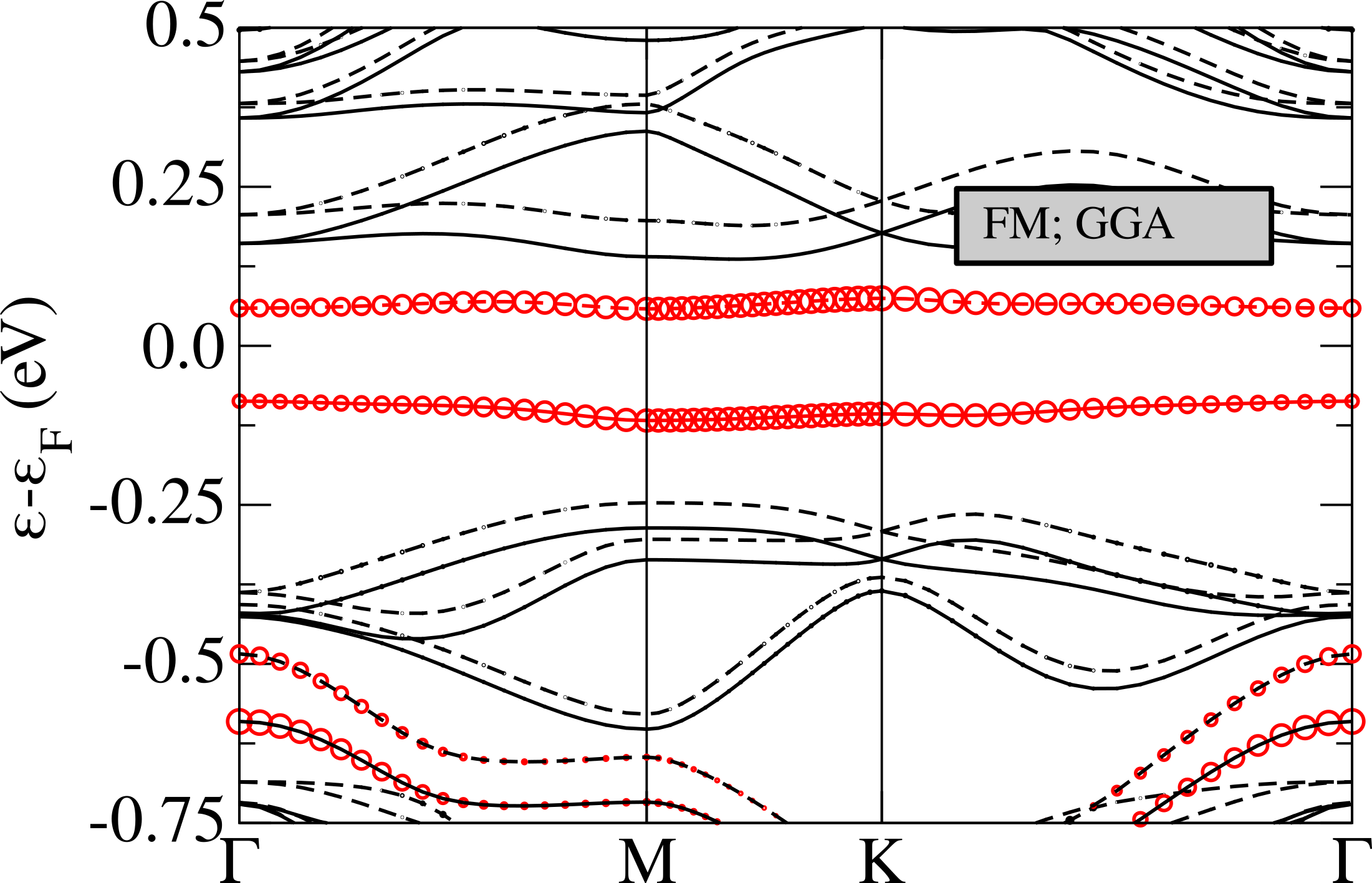}\quad\includegraphics[width=\columnwidth]{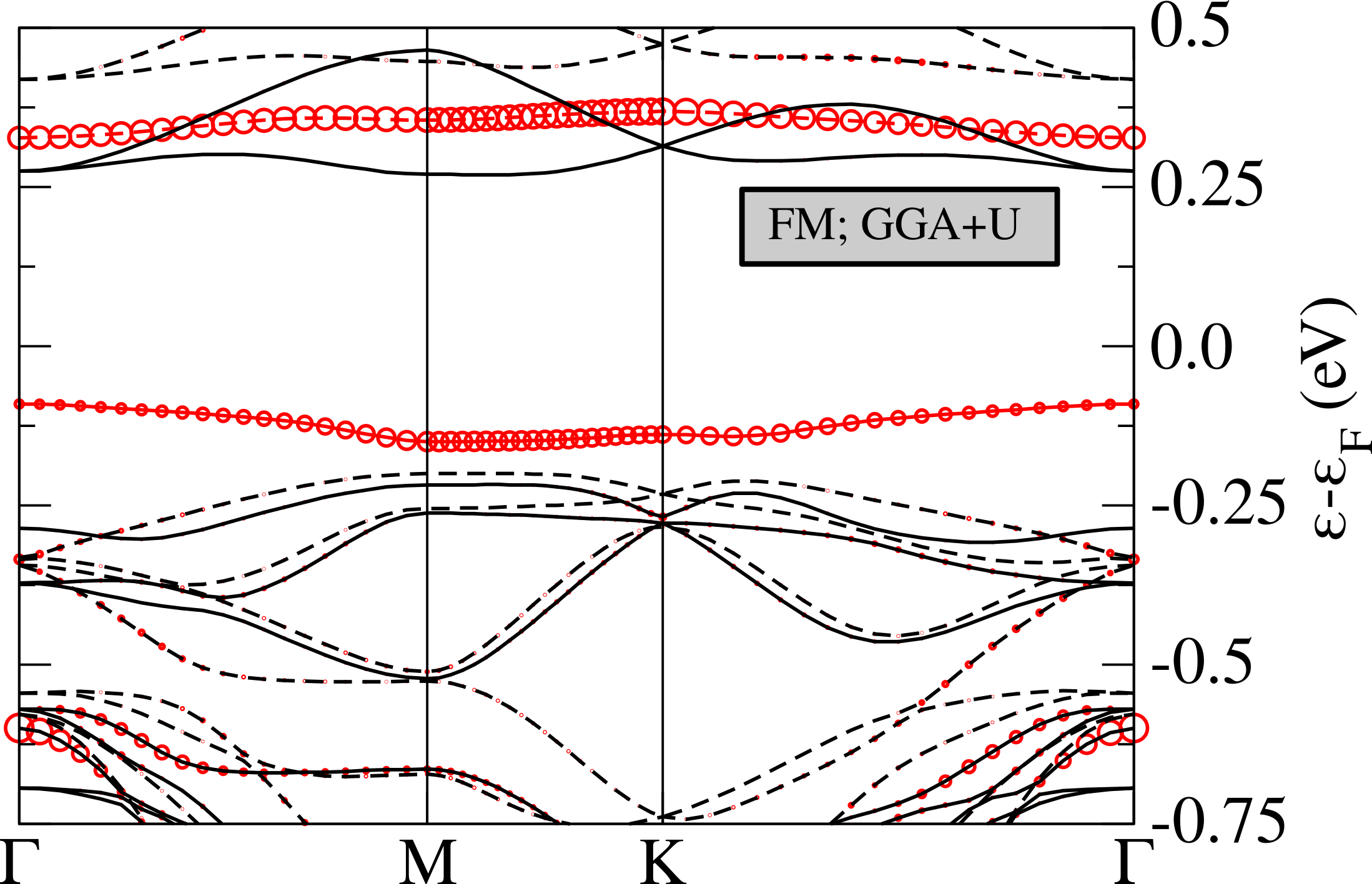}
\caption{Spin resolved electronic band structures for the $\sqrt{13}\times\sqrt{13}$ phase obtained in GGA (left panel) and in GGA+U approximation (right panel); solid (dotted) lines are related to majority (minority) spin states. The size of the circles is proportional to the central Nb $d_{z^2-r^2}$ character of the eigenvalues (the percentage of the central Nb $d_{z^2}$ component for the flats bands at $\Gamma$ are 9.5\% for the majority (12\% for the minority) component in the GGA approximation and 6.2\% for the majority (16\% for the minority) component in the GGA+U one). }\label{fig6}
\end{figure*}

The magnetic bands are reported in Fig.\ref{fig6}~(left), as we can
see, in the ferrimagnetic configuration, the system results to be semiconductor with a gap of about
0.15~eV. The flat band is splitted in two, one is fully occupied and
the second one is empty, so that the total magnetic moment is
 1~$\mu_B$. By referring 
to the left panel in Fig.\ref{fig4}, the magnetic moments are
0.21~$\mu_B$ on the red sites, 0.04~$\mu_B$ on the blue ones and 
negligible contributions on the others. 
It is worth to underline that the magnetic structure is still unstable (4.4~mRy/Nb) with respect to the 1H one.

The insurgence of magnetism induces a weak hardening of some
A$_{2g}$ modes,
in principle detectable as a Raman shift (see App.~B).

However, given the correlated nature of the problem and the key
role of the DFT+U approximation in determining total energies, as
shown in 1T-NbSe$_2$\cite{PhysRevLett.121.026401},
we perform DFT+U calculations using the method in Ref.~\cite{PhysRevB.71.035105}. The $U$ parameter is computed 
self-consistently from first principles\cite{PhysRevB.71.035105}, we obtain $U=2.87$~eV. This
value is similar to those found for 1T-TaS$_2$\cite{PhysRevB.90.045134} and
1T-NbSe$_2$\cite{PhysRevLett.121.026401} compounds.

We first perform structural optimization of the high-symmetry
1T-NbS$_2$ and 1H-NbS$_2$ single-layer structure within DFT+U.
For the 1H-polytype, we find that the in-plane lattice
parameter in DFT+U is in slightly better agreement with the one measured in the
bulk than in the GGA case (see Tab.~\ref{tab1}),
suggesting that DFT+U gives a slightly better energetic then GGA, as
it happens in NbSe$_2$\cite{PhysRevLett.121.026401}.
As it can be seen, the energy difference between the 1H and 1T
undistorted polytypes is now reduced. 

We then optimize the geometry in the CDW phase with DFT+U obtaining a small
contraction of the in-plane lattice constants (see  Tab.~\ref{tab1}, the Wyckoff positions are reported in App.~A). 
Also the magnetic structure is slightly different from GGA results: a stronger
ferrimagnetic solution with magnetic moments of 0.41~$\mu_B$ on the
red sites, 0.03~$\mu_B$ on the blue ones and negligible
antiferromagnetic contributions from the other sites is stabilized (we refer to the
left panel in Fig.\ref{fig4} for color labeling). The total magnetic moment per unit cell is of
1~$\mu_B$. The magnetic moment on the central atom is thus almost
the double of the one found in spin-polarized GGA (while the total
spin is of course still $1/2$).
Moreover, as shown in Tab.~\ref{tab1}, the CDW magnetic
solution has an  
important energy gain with respect to the 1T-polytype and it is almost
degenerate with the 1H one (0.5 mRy energy difference).

This energy difference is slightly smaller than the one found between
1H-NbSe$_2$ highly symmetric polytype and the 1T-NbSe$_2$ charge
density wave phase. This suggest that 1T-NbS$_2$ can be synthesized with
a similar experimental procedure to the one used for NbSe$_2$
\cite{Nakata2016,doi:10.1021/acs.chemmater.7b03061}.

In Fig.\ref{fig6} we report the band structure for the magnetic
solution obtained in DFT+U compared with the ones 
obtained in GGA. The band-gap between the flat band is now more than
twice that in spin-polarized GGA ($\sim$0.41~eV) and is not the
fundamental gap as the minority spin flat bands is pushed inside the empty $d$-conduction bands. The fundamental gap is $\sim$0.35~eV and involves
transitions between $d$-orbitals of different Nb atoms.

Finally we evaluate the nearest-neighbor ($J_1$) and
next-near-neighbor ($J_2$) exchange constants between different star
of David clusters in an ferromagnetic Hubbard model described by the
following
Hamiltonian: $$\hat{\mathcal{H}}=-\frac{J_1}{2}\sum_{<i,j>}\hat{S_i^z}\cdot
\hat{S_j^z} -\frac{J_2}{2}\sum_{<<i,j>>}\hat{S_i^z}\cdot
\hat{S_j^z}.$$
We adopted a super-cell approach and considered a $(2\sqrt{13})\times(3\sqrt{13})$ cell with different collinear magnetic configurations. It is important to note that we obtain similar total energies for all the spin configurations took into account. 

The calculated ferromagnetic exchange couplings are $J_1$=9.5~K and $J_2$=0.4~K, in line with the parameters describing the similar NbSe$_2$ compound\cite{Pasquier1TNbS2}. From that the system results to have a ferromagnetic ground state between different stars.

\section{Conclusions}

In this work we investigated by first principles the possible
formation of single-layer 1T-NbS$_2$ as well as its structural,
electronic and dynamical properties 
in the high symmetry phase and in the CDW one with different degrees
of correlation and allowing for magnetic solutions. 

We demonstrate that the 1T undistorted ($1\times 1$) polytype is
highly unstable towards a $\sqrt{13}\times\sqrt{13}$ reconstruction.
Within the GGA+U approximation, the $\sqrt{13}\times\sqrt{13}$
structural distortion and the
formation of a ferrimagnetic state cooperate in stabilizing the 
1T-NbS$_2$ phase in single layer form that becomes comparable
in energy with that of the 1H polytype.  Thus, we predict that 
this system can be synthesized with similar techniques to those
used for single layer
1T-NbSe$_2$\cite{Nakata2016,doi:10.1021/acs.chemmater.7b03061}.
Interestingly, a previous work\cite{B315782M} describes growth of bulk
1T-NbS$_2$ on glass keeping the substrate at very high temperature.
As a similar technique has been used for the synthesis of 1T-NbSe$_2$
(with a different
substrate) in Ref. ~\cite{Nakata2016}, this makes the synthesis of 
single layer 1T-NbS$_2$ even more likely.
 
Finally, it is interesting to underline that in this system,
magnetism occurs in  a ultraflat band, isolated from all the others
and having a marked d$_{z^2-r^2}$ character on the central Nb atom in
the star.  Spin polarized calculations without any Hubbard mean field
term, recover the insulating state by stabilizing magnetism, (although with a fairly small gap). 
A similar effect occurs in TaS$_2$\cite{PhysRevB.90.045134}, where even at $U=0$ a magnetic state is
stabilized within the spin polarized generalized gradient
approximation.
In this respect, sulfides are 
odd with 1T-NbSe$_2$ where the flat band is strongly hybridized with Se
states and the Hubbard interaction is needed to disentangle it from
the other bands\cite{PhysRevLett.121.026401}.
1T-NbS$_2$ in the $\sqrt{13}\times\sqrt{13}$ is then a prototype system
where the presence of an ultraflat band produces magnetism even at
very moderate values of $U/t$.

\section*{Acknowledgments} 
We thank G. Menichetti for useful and stimulating discussions.

This work was supported by French state funds managed by the ANR
within the Investissements d'Avenir program under references ANR-13-IS10-0003-01
ANR-11-IDEX-0004-02,  and more specifically within the framework of the
Cluster of Excellence MATISSE led by Sorbonne Universit\`e, by the
European Graphene Flagship (GrapheneCore 2).
Computer facilities were provided by CINES, IDRIS, and CEA
TGCC and PRACE (2017174186).

\section*{Appendix A}

We report the relaxed Wyckoff positions for the low temperature $\sqrt{13}\times\sqrt{13}$ CDW (non-magnetic) phase.
Both structures obtained in GGA and GGA+U belongs to the $P\bar{3}$ space group (group number 147), and the Wyckoff positions are reported in Tab. \ref{WP1} and Tab. \ref{WP2} respectively.

\begin{table}[h!]
\begin{center}
\begin{footnotesize}
\begin{tabular}{ c c c c c c }
\hline
\hline
 C=(0,0,0)    & multiplicity  &  Wyckoff label     & x & y & z   \\
\hline
Nb & 1 & a & 0.00000  & 0.00000 & 0.00000 \\ 
Nb & 6 & g & 0.28854  & 0.07046 & 0.00044  \\ 
Nb & 6 & g & 0.63662  & 0.15249 & -0.00124 \\
S  & 6 & g & 0.05083  & 0.17491 & 0.13591  \\ 
S  & 6 & g & 0.35363  & 0.25132 & 0.13360 \\ 
S  & 6 & g & 0.48561  & 0.19908 & 0.87839  \\
S  & 2 & d & 0.33333  & 0.66667 & 0.88017  \\   
S  & 6 & g & -0.02692 & 0.40822 & 0.11972  \\
\hline
\hline
\end{tabular}
\end{footnotesize}
\end{center}
\caption{Wyckoff positions of the 1T-NbS$_2$ $\sqrt{13}\times\sqrt{13}$ CDW phase obtained in GGA.}
\label{WP1}
\end{table}

\begin{table}[h!]
\begin{center}
\begin{footnotesize}
\begin{tabular}{ c c c c c c }
\hline
\hline
 C=(0,0,0)   & multiplicity   &  Wyckoff label          & x & y & z   \\
\hline
Nb & 1 & a & 0.00000  & 0.00000 & 0.00000 \\ 
Nb & 6 & g & 0.28891  & 0.07059 & 0.00036  \\ 
Nb & 6 & g & 0.63634  & 0.15237 & -0.00097 \\
S  & 6 & g & 0.05072  & 0.17451 & 0.13662  \\ 
S  & 6 & g & 0.35369  & 0.25129 & 0.13419 \\ 
S  & 6 & g & 0.48561  & 0.19901 & 0.87758  \\
S  & 2 & d & 0.33333  & 0.66667 & 0.87948  \\   
S  & 6 & g & -0.02695 & 0.40829 & 0.12062 \\
\hline
\hline
\end{tabular}
\end{footnotesize}
\end{center}
\caption{Wyckoff positions of the 1T-NbS$_2$ $\sqrt{13}\times\sqrt{13}$ CDW phase obtained in GGA+U (U=2.87~eV).}
\label{WP2}
\end{table}

\section*{Appendix B}

\begin{figure}[]
\centering
\includegraphics[width=\columnwidth]{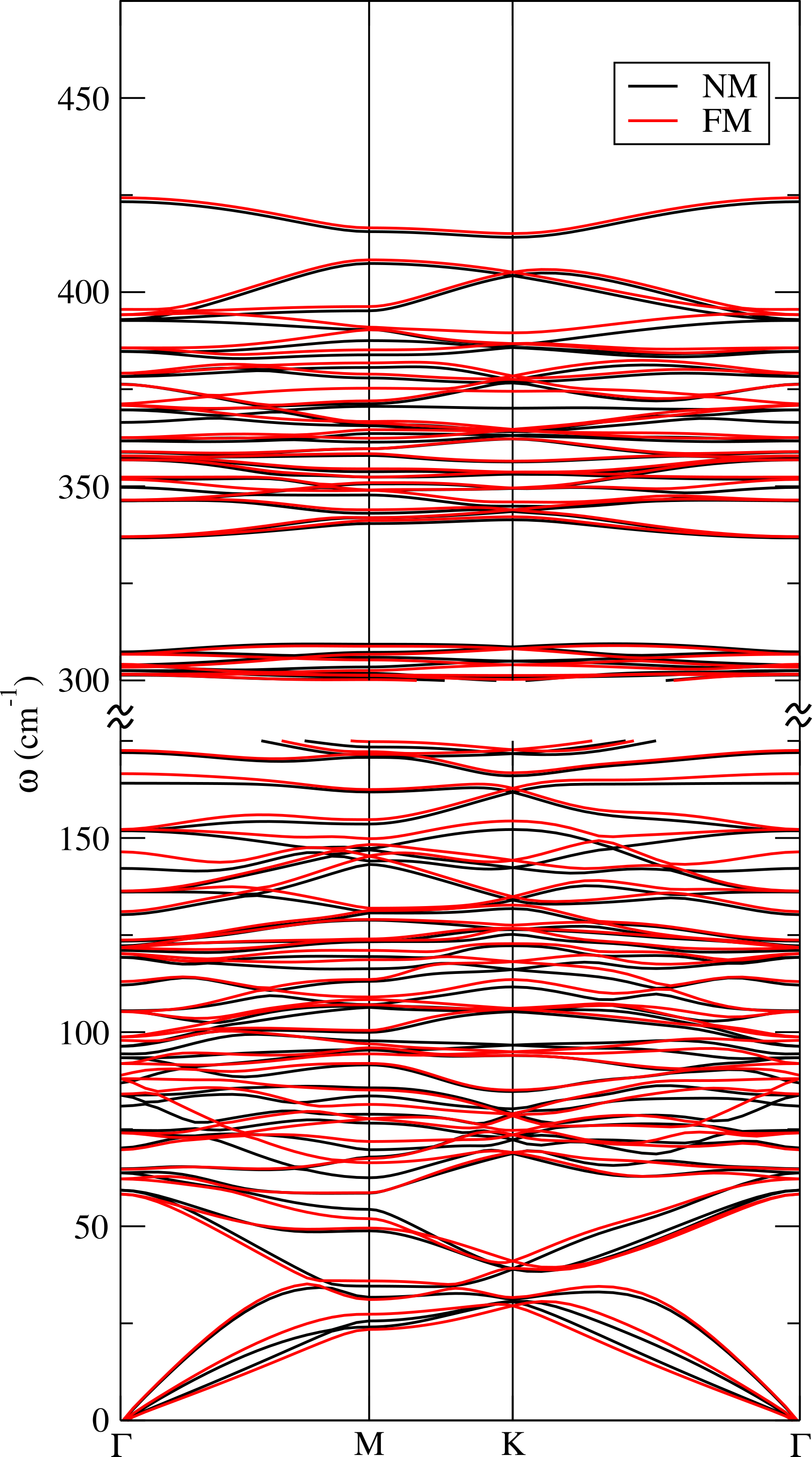}
\caption{Phonon dispersion for the $\sqrt{13}\times\sqrt{13}$ CDW phase in the non magnetic solution (black curves) and in the ferrimagnetic one (red curves).}\label{fig7}
\end{figure}

We compute the phonon dispersion for the $\sqrt{13}\times\sqrt{13}$
CDW phase in GGA approximation for the non magnetic (NM) and the
ferrimagnetic (FM) solutions. We calculate the dynamical matrix at
zone center and then Fourier interpolate the dynamical matrices on the full
Brillouin zone. Results are reported in
Fig.\ref{fig7}. All frequencies are positive revealing the CDW phase
is dynamically stable. The Raman active modes are reported in
Tab. \ref{R_ph}. The insurgence of magnetism causes a weak hardening
of A$_{2g}$ frequencies at around 90~and~366~cm$^{-1}$ (we register
shifts of about $4\div5$~cm$^{-1}$, see  Tab. \ref{R_ph}).

\begin{table}[h!]
\begin{center}
\begin{tabular}{ l c c  }
\hline
\hline
Point  & $\omega_{NM}$    & $\omega_{FM}$ \\
group   & (cm$^{-1}$)   & (cm$^{-1}$) \\
\hline
E$_g$  &   63.9  &      61.6   \\
A$_g$  &   64.6  &      64.6  \\
A$_g$  &   82.1  &      83.2   \\
E$_g$  &   84.4  &      88.6  \\
E$_g$  &   93.3  &      91.5  \\
A$_g$  &   94.2  &      97.9  \\
E$_g$  &  106.0  &     105.4  \\
E$_g$  &  119.5  &     120.2  \\
A$_g$  &  121.2  &     121.7  \\
A$_g$  &  142.9  &     146.3  \\
A$_g$  &  164.0  &     166.7  \\
E$_g$  &  179.7  &     180.4  \\
A$_g$  &  211.1  &     211.1  \\
E$_g$  &  218.8  &     218.8  \\
A$_g$  &  221.3  &     221.6  \\
E$_g$  &  226.9  &     227.1  \\
E$_g$  &  236.2  &     236.3  \\
E$_g$  &  242.2  &     242.4  \\
A$_g$  &  243.7  &     243.6  \\
E$_g$  &  257.7  &     258.6  \\
E$_g$  &  261.6  &     261.7  \\
A$_g$  &  262.3  &     262.7  \\
A$_g$  &  266.8  &     266.7  \\
E$_g$  &  268.1  &     268.6  \\
E$_g$  &  276.7  &     276.9  \\
A$_g$  &  283.9  &     283.6  \\
A$_g$  &  286.0  &     285.5  \\
A$_g$  &  298.7  &     298.7  \\
A$_g$  &  303.4  &     303.2  \\
E$_g$  &  307.5  &     306.9  \\
E$_g$  &  346.5  &     346.5  \\
A$_g$  &  349.9  &     351.7  \\
E$_g$  &  357.5  &     357.9  \\
A$_g$  &  358.9  &     359.1  \\
A$_g$  &  366.7  &     371.0  \\
E$_g$  &  378.4  &     379.0  \\
A$_g$  &  392.8  &     394.1  \\
E$_g$  &  393.0  &     395.1 \\
\hline
\hline
\end{tabular}
\end{center}
\caption{Raman active frequencies for the 1T-NbS$_2$ $\sqrt{13}\times\sqrt{13}$ CDW phase obtained in non magnetic (NM) and Ferrimagnetic (FM) phases (in GGA approximation).}
\label{R_ph}
\end{table}

\vspace{10mm}

\bibliography{bibliography}{}

\end{document}